\PassOptionsToPackage{table}{xcolor}

\documentclass[sigconf]{acmart}
\usepackage{comment}
\usepackage{tabularx}
\usepackage[table]{xcolor}
\usepackage{multirow}
\usepackage[absolute]{textpos}

\renewcommand\footnotetextcopyrightpermission[1]{}

\AtBeginDocument{%
	}

\copyrightyear{2024}
\acmYear{2024}
\setcopyright{acmlicensed}\acmConference[RAIE '24]{2024 International
	Workshop on Responsible AI Engineering }{April 16, 2024}{Lisbon, Portugal}
\acmBooktitle{2024 International Workshop on Responsible AI Engineering
	(RAIE '24), April 16, 2024, Lisbon, Portugal}
\acmDOI{10.1145/3643691.3648588}
\acmISBN{979-8-4007-0572-4/24/04}

\begin{document}
	
	\title{Do Generative AI Tools Ensure Green Code? An Investigative Study}

	\author{Samarth Sikand$^\dagger$, Rohit Mehra$^\dagger$, Vibhu Saujanya Sharma$^\dagger$, Vikrant Kaulgud$^\dagger$, Sanjay Podder$^\ddagger$, Adam P. Burden*} 
	\affiliation{ 
		\institution{$^\dagger$Accenture Labs, India \hspace{0.4em}
			$^\ddagger$Accenture, India \hspace{0.4em}
			*Accenture, USA}
		\country{}
	}
	\email{{{samarth.sikand, rohit.a.mehra, vibhu.sharma, vikrant.kaulgud, sanjay.podder, adam.p.burden}@accenture.com}}
	
	\renewcommand{\shortauthors}{Sikand et al.}
	
	\begin{abstract}
	
		Software sustainability is emerging as a primary concern, aiming to optimize resource utilization, minimize environmental impact, and promote a greener, more resilient digital ecosystem. The sustainability or 'greenness' of software is typically determined by the adoption of sustainable coding practices. With a maturing ecosystem around generative AI, many software developers now rely on these tools to generate code using natural language prompts. Despite their potential advantages, there is a significant lack of studies on the sustainability aspects of AI-generated code. Specifically, how environmentally friendly is the AI-generated code based upon its adoption of sustainable coding practices? In this paper, we present the results of an early investigation into the sustainability aspects of AI-generated code across three popular generative AI tools — ChatGPT, BARD, and Copilot. The results highlight the \textit{default non-green} behavior of tools for generating code, across multiple rules and scenarios. It underscores the need for further in-depth investigations and effective remediation strategies.
	\end{abstract}

	\ccsdesc[500]{Social and professional topics~Sustainability}
	\ccsdesc[500]{Computing methodologies~Natural language generation}
	

	\maketitle
	\vspace{-0.25em}
	
	\begin{textblock*}{10cm}(1.5cm,26.2cm) 
		DOI: \url{https://doi.org/10.1145/3643691.3648588}
	\end{textblock*}
	
	\section{Introduction}
	
	Despite playing a pivotal role in advancing sustainability across various domains, software systems exert an often underestimated carbon footprint, thereby emerging as a significant and rapidly evolving contributor to global carbon emissions. Multiple studies have estimated that the internet and communications technology industry, which encompasses software and the corresponding hardware, currently accounts for 2-7\% of global greenhouse gas emissions and is predicted to increase to a massive 14\% by 2040 \cite{unepccc, FREITAG2021100340}. One of the primary reasons behind these high carbon emissions is the non-optimization of software code from a sustainability perspective, specifically green, energy, and emissions perspective \cite{10.1145/3154384}. For example, usage of energy-hungry design patterns in code, as opposed to energy-efficient design patterns \cite{6224257}. Among other reasons, this non-optimization can majorly be attributed to the lack of awareness on the part of the developer, leading to lower adoption of sustainable coding practices \cite{9793921, 10.1145/3350768.3350770}. Lower adoption of sustainable coding practices will lead to unoptimized code, eventually leading to higher energy consumption and carbon emissions. Moreover, in traditional human-based development, the primary responsibility lies with the developer to incorporate these sustainable best practices into the code and optimize it accordingly.
	
	With rapid advancements in generative AI, particularly in AI-assisted coding, numerous software developers are already utilizing or considering the integration of these tools into their daily coding activities \cite{github}. These tools include specialized code generation/completion tools like GitHub Copilot \cite{github-copilot}, Tabnine \cite{tabnine}, and others, as well as generic large-language models (LLMs) with code generation capabilities-based tools, such as OpenAI ChatGPT \cite{chatgpt}, Google BARD \cite{bard}, Meta Code Llama \cite{llama}, among others. Research indicates that developers utilize these tools to expedite task completion, reduce code searches, and enhance code quality, among other benefits \cite{10.1145/3491101.3519665, 11223344, github}. This results in an overall increase in developer productivity and software quality. As a result of the potential significant benefits offered by these tools, their adoption will rapidly surge in the near future. Estimates suggest that by 2027, approximately 70\% of all professional software developers will be utilizing AI-assisted coding tools for their day-to-day coding-related activities, leading to a majority of the future software code being AI-generated \cite{gartner}. Additionally, in this modern human-AI teaming-based software development, the burden of ensuring the adoption of sustainable coding practices gets distributed between the human developer and the AI, with the latter playing a more proactive role in integrating and promoting environmentally friendly coding approaches.
	
	While code generation tools have been rapidly adopted and offer numerous advantages, there is a notable lack of evaluation studies focusing on the sustainability aspects of the generated code. Specifically, there is a gap in understanding how sustainable or environmentally friendly the code produced by these tools is, based on the implementation of sustainable coding practices. Although AI-generated code has been recently studied for other critical software engineering aspects such as security, performance, correctness, quality, and maintainability, etc. sustainability has been largely overlooked \cite{Du2023ClassEvalAM, khoury2023secure, Yetistiren2023EvaluatingTC}. Conducting evaluation studies in this regard can provide insights into the sustainability efficacy related to the usage of these tools, thereby influencing their adoption or non-adoption from a sustainability perspective. Furthermore, these evaluations may extend to the development of approaches and tools to address the issue in various scenarios where it may arise.
	In this paper, we present an early exploration aimed at studying the sustainability aspects of AI-generated code, grounded in the adoption of sustainable coding practices. The study encompasses three widely used AI code generation tools: ChatGPT, BARD, and Copilot. These tools were assessed for their adherence to six sustainable coding practices, selected from previous sustainability research and standard knowledge bases. Preliminary results highlight the non-green default behavior of the evaluated tools across multiple scenarios. If adopted \textit{as-is} during software development, the generated non-green code could contribute to excess energy consumption and carbon emissions, thereby negatively impacting the sustainability of our environment.
	\begin{figure}[t]
		\centering
		\includegraphics[width=0.6\linewidth]{"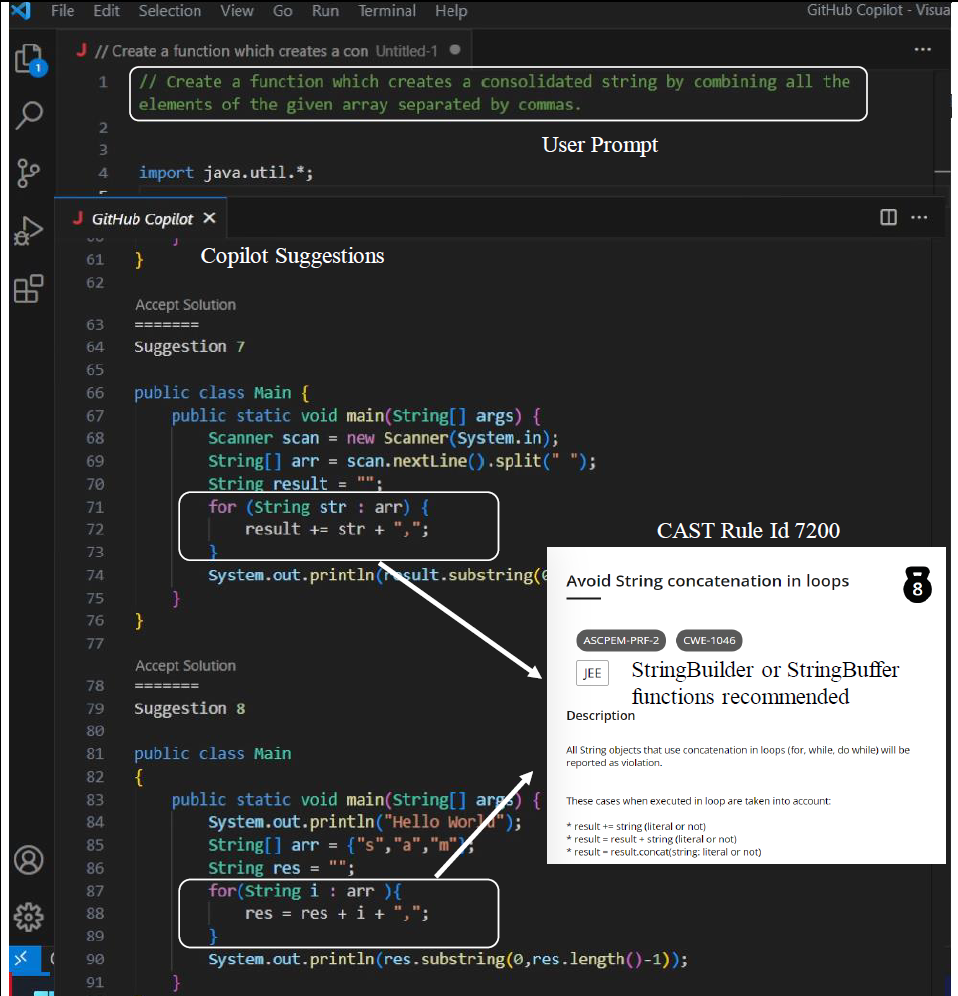"}
		\caption{Illustration of ``default'' behavior of Gen AI in generating \textit{energy-inefficient} outputs.}
		\label{fig:sample_issue}
	\end{figure}

	\section{Evaluating The Greenness of AI Generated Code}
	To evaluate the GenAI tool's adherence to sustainable practices, we formulated an effective methodology that will help understand the tool's adherence to those practices. To that end, we create a \textit{nano-dataset} of prompts, for whom sustainable a priori patterns/solutions would be known. The presence or absence of green/sustainable patterns in tool outputs would inform our understanding of the tool's default behavior. In further subsections, we will delve deeper into the methodology and discuss the outcomes of our study.
	
	\subsection{Study Methodology}
\begin{figure}[t]
		\centering
		\includegraphics[width=\linewidth]{"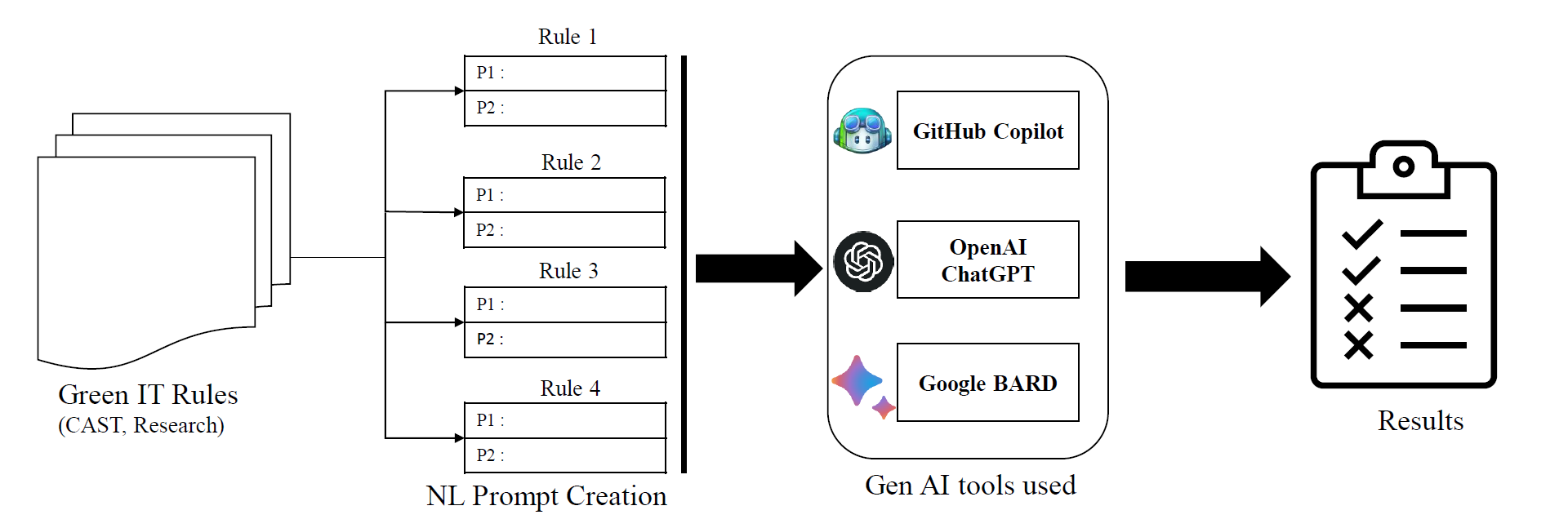"}
		\caption{Overall approach to investigative study.}
		\label{fig:approach}
	\end{figure}

	Fig \ref{fig:approach} illustrates our overall study approach. Our approach has approximately three high-level steps : (i) Choosing an appropriate set of green rules. (ii) Shortlisting Gen AI tools to be evaluated (iii) NL Prompt creation. To narrow the scope of the study, the prompts used for evaluation were created based on the context of the rules rather than framing the prompts independently. The main reason for this design choice was to influence the LLM behavior to generate specific outputs pertaining to the rule’s context by solving a directed task
	
	\subsubsection{Choosing Green Rules}
	
	While many research studies have been conducted to find energy-hungry code patterns in code, there is a lack of a standardized dataset or knowledge base for such patterns. One of the more known rule sets is CAST’s Green IT rules\cite{cast01}, which enumerates many energy-inefficient code patterns. The Green IT rules comprise a set of energy-efficient coding practices that can be segregated by technologies(JEE, Python, SAP, etc.), criticality, and other categorizations. Currently, there exist approximately 900 rules in the Green IT set of rules, from which we considered rules relevant for Java, Javascript and Python language only. 

	For the purpose of this study, we chose 5 rules from CAST Green IT rules and 1 rule was constructed from a study evaluating the energy footprint of various Java I/O APIs \cite{rocha2019comprehending}. Table 1 demonstrates the rules chosen in our study. The underlying reason for choosing these rules is twofold (i) easy to validate the code-snippet/source code manually (ii) the underlying fundamental task(e.g., I/O operations, loops) is relatively easy to replicate.
	
	\subsubsection{Shortlisted Generative AI tools}
	
		\begin{table*}[t]
		\caption{Prompts crafted for each of the rules and the evaluation results of each tool on respective prompts. \textit{\textcolor{red}{Red}} indicates absence of green patterns in any solution, \textit{\textcolor{green}{Green}} indicates presence of green patterns in all presented solutions and \textit{\textcolor{orange}{Orange}} indicates partial presence of green patterns(provided with \%age. \%age is computed by \{\# of \textit{green} suggestion/total suggestions shown by Copilot\})}
		\label{tab:results}
		\small
		\centering
		\begin{tabular}{|>{\centering\arraybackslash}p{3em}|>{\centering\arraybackslash}m{3cm}|>{\centering\arraybackslash}p{2.5cm}|>{\raggedright\arraybackslash}m{6cm}|>{\centering\arraybackslash}m{3em}|>{\centering\arraybackslash}m{4em}|>{\centering\arraybackslash}m{3em}|}
			\hline
			&&&&&&\\[-1em]
			Rule \# &Rule Title & Description & Prompt & Copilot & ChatGPT & BARD\\
			\hline
			&&&&&&\\[-1em]
			\multirow{2}{3em}[-1em]{Rule 1} & \multirow{2}{3cm}[-1em]{Comprehending Energy Behaviors of Java I/O APIs} & \multirow{2}{2.5cm}[0.5em]{I/O APIs like \texttt{Scanner} and \textit{FileInputStream} are not energy efficient} & Create a class which reads a 20MB HTML file and prints the number of words in the file & \cellcolor{red}& \cellcolor{red} &\cellcolor{red}\\ 
			
			\cline{4-7}
			&&&&&&\\[-1em]
			& & & Create a function which takes text file as input and outputs the number of words of file & \cellcolor{red}& \cellcolor{red} &\cellcolor{red}\\
			\hline
			
			&&&&&&\\[-1em]
			\multirow{2}{3em}[-1em]{Rule 2} & \multirow{2}{3cm}[-1em]{Avoid String concatenation in loops} & \multirow{2}{2.5cm}[1.3em]{Create \textit{StringBuilder} or \textit{StringBuffer} before entering loop, and append to it within loop}& Create a function which creates a consolidated string by combining all the elements of the given array separated by commas.& \cellcolor{red}& \cellcolor{green} &\cellcolor{green}\\ 
			
			\cline{4-7}
			&&&&&&\\[-1em]
			& & & Create a function to parse a JSON file and output all the data in a consolidated string & \cellcolor{red}& \cellcolor{green} &\cellcolor{green}\\
			\hline
			
			&&&&&&\\[-1em]
			\multirow{2}{3em}[-1em]{Rule 3} & \multirow{2}{3cm}[-1em]{Use varchar2 instead of char and varchar} &\multirow{2}{2.5cm}{Change char and VARCHAR columns to VARCHAR2}& Write a SQL query for creating a table for Pets. The columns should store about information a Pet's unique ID, name, age, color, breed, weight, owner's ID, and owner's name.& \cellcolor{red}& \cellcolor{red} &\cellcolor{red}\\ 
			
			\cline{4-7}
			&&&&&&\\[-1em]
			& & & Write a SQL query to create tables for the following Java class: <<\texttt{code}>> & \cellcolor{red}& \cellcolor{red} &\cellcolor{red}\\
			\hline
			
			&&&&&&\\[-1em]
			\multirow{2}{3em}[-0.5em]{Rule 4} & \multirow{2}{3cm}[-0.5em]{Avoid using HashTable
			} & \multirow{2}{2.5cm}[0em]{\textit{HashMap} should be preferred over \textit{HashTable}}& Write a java function to print duplicate entries in an array.& \cellcolor{green}& \cellcolor{green} &\cellcolor{red}\\ 
			
			\cline{4-7}
			&&&&&&\\[-1em]
			& & & Write a java function to create a lookup table of locally stored username and passwords. & \cellcolor{green}& \cellcolor{green} &\cellcolor{green}\\
			\hline
			
			&&&&&&\\[-1em]
			\multirow{2}{3em}[-0.5em]{Rule 5} & \multirow{2}{3cm}[-0.5em]{Avoid using forEach()
			} & \multirow{2}{2.5cm}[0.5em]{Function-based iteration takes up to eight times as long as loop-based iteration.} & Write a javascript function to print all values of an array.	& \cellcolor{orange}50\% (5/10) & \cellcolor{green} &\cellcolor{green}\\ 
			
			\cline{4-7}
			&&&&&\\[-1em]
			& & & Write a javascript function to print all entries of a string array that are palindrome. & \cellcolor{orange}60\% (6/10) & \cellcolor{green} &\cellcolor{green}\\
			\hline
			
			&&&&&&\\[-1em]
			\multirow{2}{3em}[-1em]{Rule 6} & \multirow{2}{3cm}[-1em]{Avoid leaving open file resources (Python)} &\multirow{2}{2.5cm}{Open file using \textit{with} statement or explicitly close opened files.}& Write a python function to print the contents of a local file.& \cellcolor{green}& \cellcolor{green} &\cellcolor{green}\\ 
			
			\cline{4-7}
			&&&&&&\\[-1em]
			& & & Write a python function to prepend a line "Accenture-Proprietary" at the start of each file. Filenames are listed in an array & \cellcolor{orange}62.5\% (5/8)& \cellcolor{green} &\cellcolor{green}\\
			\hline
		\end{tabular}
	\end{table*}
	
	Since ChatGPT’s inception, there has been a meteoric rise in foundational models, like LLMs, as every organization is constantly striving to push state-of-the-art performance with its models. For our study, we chose 3 popular Generative AI tools with thousands(or millions) of users, which are as follows:
	\begin{itemize}
		\item{\textbf{Github Copilot}} : Copilot, released in 2021, has become one of the most popular AI-powered code generation tools today. Recently, Github claimed to have more than a million developers using their tool\cite{github01}.
		\item{\textbf{OpenAI ChatGPT}}: ChatGPT has become one of the fastest-growing internet tools ever, by amassing more than 100 million users in a span of 2 months\cite{verge02}. One of the most popular AI tools in recent times and its near-human level performance makes it a prime candidate for evaluation. 
		\item{\textbf{Google BARD}}: Google released BARD in early 2023\cite{google01}, which claimed to overcome some of the shortcomings of ChatGPT, like pulling real-time information from the World Wide Web.
	\end{itemize}
	
	The tools chosen reflect the diversity in terms of approaches taken to train the models and how they are packaged to be used. We expect that results from these tools will inform our understanding of these model’s knowledge and reasoning capabilities regarding energy-efficient coding practices.
		
	\subsubsection{Prompt creation}
	
	For testing the efficacy of Gen AI tool’s adherence to Green best practices (without external/additional inputs), we manually created two Natural Language (NL) prompts for each of the six rules shortlisted. The NL prompts, for each rule, are crafted in a manner to ensure that the underlying task of the respective rule will be implemented. The outputs of the Gen AI tools will represent their default behavior on the prompt tasks. Table \ref{tab:results} describes all the prompts crafted for the chosen rules

	\subsection{Experiment Results and Discussion}
	
	To conduct our experiments, we leveraged ChatGPT and BARD conversational UI tools on their respective organization's sites. Additionally, for rules pertinent to Java/JEE we prepend each NL prompt with the following statement: ``\texttt{All responses should be for Java language}'', to warrant the usage of Java syntax in tool's outputs. For Copilot, the official VSCode extension was used to generate up to 10 solutions(default setting) for a NL prompt, by explicitly triggering Copilot suggestions panel. In Table \ref{tab:results}, we present our findings for each combination of rule and GenAI tool. It can be clearly inferred that at least one GenAI tool is not following the chosen Green IT rules. For Rule 1, none of the tools provided the most optimum solution. The I/O functionality is one of the most fundamental of tasks and it can have a significant energy impact, as it can have a multiplicative effect of being used in millions of developer codebases. Gen AI's inclination towards Java I/O functions, like \texttt{Scanner} and \texttt{FileInputStream}, can be attributed to the popularity and ease of use of such I/O functions in majority of codebases. We also observe that most suggestions by Github Copilot do not exhibit \textit{Green} coding patterns, for the rules under study. For Rule 5 and Rule 6, at least 40\% of the suggestions from Copilot are non-green. Meanwhile, ChatGPT fails to adhere to sustainable practices for 33\% of the selected rules. BARD follows closely behind ChatGPT in terms of non-adherence to rules in its solutions. 
	
	It can be observed from the results that even the basic or fundamental sustainable practices are not followed consistently across different tools. We have observed Copilot recommended solutions to be less sustainable than that of ChatGPT or BARD, as at least one of its solutions being non-green for 5 out of 6 rules. These observations can be attributed to multiple factors, like differences in training data and training methods for underlying models of Copilot, ChatGPT, and BARD. Although the results present an irrefutable fact that GenAI tools may not provide sustainable responses for every scenario, we believe the tool's responses can be improved by certain actions. 
	
	Depending on the access to underlying LLM model of GenAI tool, one can mitigate these sustainability shortcomings. If an organization has partial/full access to underlying model, they can \textit{fine-tune the LLMs} to include knowledge around sustainable coding practices. But if they leverage Gen AI tool through blackbox APIs, they can leverage \textit{Prompt engineering} methods to explicitly append relevant sustainable coding practice to a prompt. The practical solution for anyone would depend on business use case and other constraints (e.g., budget/compute constraints).
	
	\section{Limitations}
	
	While we initiated this study to investigate the default behavior of Generative AI tools to generate green software artifacts, it was not feasible to cover all the energy-efficiency coding patterns and best practices. In one of the initial experiments, we discovered that many of the rules descriptions were difficult to replicate with a NL prompt. For example, an empirical study put forth that Inheritance is more energy-efficient than Delegation pattern\cite{connolly2020inheritance}. We attempted to craft a NL prompt without explicitly mentioning the code pattern, but LLM's non-deterministic responses didn't include solutions with Inheritance or Delegation pattern. Such issues made the evaluation of many rules extremely challenging. Additionally, the small set of Green rules selected was by design, firstly due to limited run time, and secondly, to understand if this area of research was worth exploration. Hence, the current results are not representative of one tool's superior capability, to generate \textit{green} code, over another.
	
	Moreover, evaluation hundreds of LLMs\footnote{Stanford HELM : https://crfm.stanford.edu/helm/latest/\#/models}(e.g., LLAMA\cite{touvron2023llama}), from various organizations and with different licenses, would make the study extremely challenging. Due to the huge scale of effort required for exhaustive evaluation, we narrowed our scope to three tools only. Furthermore, our approach has been bottom-up i.e. we shortlisted the rules and then crafted prompts, whereas alternate approach would be to consider prompts (NL and code both) first, independent of the sustainable coding practices. While we believe that this exploratory study is a good starting point for evaluating sustainability of AI-generated artifacts, a more comprehensive and standardized approach is needed to evaluate sustainable behavior of foundation models with different modalities.

	\section{Future Work and Conclusion}
	
	In our study, we have highlighted some of the concerning “default” behavior of diverse Gen AI tools in generating green code. It can be observed that it is important to explore the default behavior of the Gen AI tools from the perspective of green code and not trust its outputs blindly. Some novel solutions can help shift some of the sustainable coding best practices considerations from the developer and reduce their rework in later stages of development along with cognitive workload. For future work, we plan to do a more exhaustive behavior profiling of a larger set of GenAI or LLM tools from the perspective of sustainability. Also, approach the problem from other way around, by setting tasks independent of the sustainable coding practices and then do the profiling. Our aim is to encourage researchers in industry and academia to evaluate the sustainability aspects of generated artifacts along with functional correctness, security, etc. 
	

\bibliographystyle{ACM-Reference-Format}
\bibliography{sample-base}

\end{document}